\documentclass[
	reprint,
	groupedaddress,
	amsmath,amssymb,
	prb,
	floatfix,
	superscriptaddress,
	column,
	aps,
]{revtex4-2}

\usepackage{amsmath}
\usepackage{bm}
\usepackage{amsfonts}
\usepackage{amssymb}
\usepackage{amsthm}
\usepackage[caption=false]{subfig}
\usepackage{graphicx}
\usepackage{hyperref}
\usepackage{cleveref}
\usepackage[T1]{fontenc}
\usepackage{libertine}
\usepackage{lineno}
\usepackage{multirow}
\usepackage{lipsum} %
\usepackage[dvipsnames]{xcolor}

\DeclareMathOperator*{\argmin}{arg\,min}

\begin{document}

\title{\textit{pyeCE}: A Python Implementation of the \textit{Embedded} Cluster Expansion}
\author{Yann L. M{\"u}ller}
\affiliation{Laboratory of materials design and simulation (MADES), Institute of Materials, \'{E}cole Polytechnique F\'{e}d\'{e}rale de Lausanne}
\affiliation{National Centre for Computational Design and Discovery of Novel Materials (MARVEL), \'{E}cole Polytechnique F\'{e}d\'{e}rale de Lausanne}
\author{Claire A. Paetsch}
\affiliation{Laboratory of materials design and simulation (MADES), Institute of Materials, \'{E}cole Polytechnique F\'{e}d\'{e}rale de Lausanne}
\affiliation{National Centre for Computational Design and Discovery of Novel Materials (MARVEL), \'{E}cole Polytechnique F\'{e}d\'{e}rale de Lausanne}
\author{Anirudh Raju Natarajan}
\email{anirudh.natarajan@epfl.ch}
\affiliation{Laboratory of materials design and simulation (MADES), Institute of Materials, \'{E}cole Polytechnique F\'{e}d\'{e}rale de Lausanne}
\affiliation{National Centre for Computational Design and Discovery of Novel Materials (MARVEL), \'{E}cole Polytechnique F\'{e}d\'{e}rale de Lausanne}

\begin{abstract}
  The cluster expansion is a widely used approach for predicting the finite-temperature thermodynamics of alloys from zero-kelvin first-principles calculations, but its conventional formulation becomes intractable for materials with more than three or four chemical species. High-entropy alloys have therefore remained largely out of reach. We present \textit{pyeCE}, an open-source Python library that implements the embedded cluster expansion (eCE), in which machine learning maps many chemical species onto a smaller set of effective species and thereby limits the growth in the number of cluster functions. \textit{pyeCE} provides the complete modeling workflow, including the construction of symmetry-adapted descriptors with a learnable per-sublattice chemical embedding, a neural-network energy model, ladder-based training, uncertainty quantification, and finite-temperature simulations through Monte Carlo sampling. Built on the \textit{PyTorch} and \textit{pymatgen} libraries, it supports systems with multiple species on multiple sublattices and runs on graphics processing units. We demonstrate the package on two material systems. In the first, a single model spanning the full composition space of a 9-component refractory alloy resolves short-range order and order--disorder behavior. This model enables rapid screening for compositions with strong Cr clustering, a feature linked to the formation of a continuous, corrosion-resistant oxide scale. In the second, a model of hydrogen dissolution in a Mo--Nb--W alloy reproduces the composition dependence of hydrogen uptake and resolves the interstitial environments that hydrogen occupies. The modular design of \textit{pyeCE} allows it to be extended to problems beyond alloy thermodynamics, including kinetics, defect energetics, and the coupling of chemical order to magnetic and vibrational degrees of freedom.
\end{abstract}

\maketitle

\section{Introduction}
\label{sec:introduction}

Predicting the finite-temperature properties of materials from zero-kelvin electronic structure calculations is an important step towards the design of structural, energy, and electronic materials, and is central to understanding phase transformations and transport mechanisms\cite{van_der_ven_first-principles_2018}. This link between the zero-kelvin and finite-temperature descriptions is commonly established through an atomistic surrogate model that reproduces the energetics of a system at a fraction of the cost of first-principles calculations. The model is then coupled with a finite-temperature sampling method such as molecular dynamics, Monte Carlo simulation, or a combination of the two together with free-energy integration techniques. Empirical interatomic potentials\cite{daw_baskes_eam}, machine-learned interatomic potentials\cite{drautz2019_ace,behlerparrinello}, and on-lattice models such as the cluster expansion\cite{sanchez1984} are all widely used for this purpose.

For alloys with a few chemical species, on-lattice cluster expansions are straightforward to parameterize, accurately reproduce ground-state energetics, and are readily sampled with Monte Carlo techniques to estimate finite-temperature properties. These features have made the cluster expansion a versatile tool across materials science. It is commonly used to construct phase diagrams and to model order--disorder transitions in alloys \cite{muller2024, van_der_ven_first-principles_2018, natarajan2017a, linderalv2022}. The formalism extends to systems with vacancies on one or more sublattices, which has led to its extensive use in modeling the intercalation of lithium, sodium, and magnesium in battery cathodes \cite{kitchaev2018, richards2018, lun2019, lun2021}, the equilibrium vacancy concentration in alloys \cite{lee2026, belak_effect_2015}, and the uptake of oxygen, carbon, nitrogen, and hydrogen in metals \cite{puchala_thermodynamics_2013, paetsch2026, gunda2020understanding,reynolds2024solute}. Coupling an energy cluster expansion with a second cluster expansion for migration barriers extends the framework to vacancy-mediated transport \cite{vanderven2001, vanderven2010,lee2026diffusion}. The framework also captures the influence of local chemical order on planar defects such as surfaces and generalized stacking faults, providing a route to predicting the defect properties that govern dislocation behavior and fracture \cite{natarajan2020a, mak2021}. Finally, it can incorporate atomistic degrees of freedom beyond chemical disorder, including vibrational and magnetic excitations and their coupling \cite{thomas2013finite, thomas2014, vandewalle2002a, kadkhodaei2017, kitchaev2021a, decolvenaere2019, drautz2004a}.

Extracting finite-temperature properties from any atomistic model requires an appropriate coarse-graining procedure. Free energies yield phase diagrams, ensemble averages of the occupations of neighboring sites quantify short-range order, and the time evolution of atomic positions gives diffusion coefficients. For on-lattice cluster expansions, the Metropolis--Hastings algorithm provides a simple basis for Monte Carlo sampling in ensembles ranging from canonical to semi-grand canonical \cite{metropolis1953, hastings_1970}. Umbrella sampling reaches the unstable regions of the free energy that serve as input to longer-length-scale methods such as phase-field modeling \cite{natarajan2017}. Diffusion is treated with rejection-free kinetic Monte Carlo built on the migration-barrier expansion. Combined with vacancy concentrations and thermodynamic factors, this approach yields the Onsager transport coefficients and, in turn, non-dilute diffusion coefficients at finite temperature\cite{vanderven2010}. A cluster expansion package must therefore supply not only the energy model but also the sampling machinery that converts it into finite-temperature properties, together with the statistical tools needed to judge when those properties have converged.

Despite this broad utility, extending the cluster expansion to alloys containing more than three or four chemical species has remained a long-standing challenge. The number of cluster functions grows polynomially with the number of species, which makes conventional cluster expansions for such systems difficult or impossible to parameterize. This limitation has kept the cluster expansion from being applied to multi-principal-element alloys, also known as high-entropy alloys, where phase stability near the center of composition space is of particular interest.

The embedded cluster expansion (eCE) formalism addresses this challenge by using machine learning to identify chemical similarities between the elements of an alloy \cite{muller2025}. An alloy with many chemical species is treated as one with fewer effective species, which substantially reduces the number of cluster functions while retaining accuracy. The embedding is a learned low-dimensional projection of the site-basis functions that is optimized jointly with the energy model, so the effective species are determined by the training data rather than chosen in advance. Established cluster expansion packages such as CASM, ICET, smol, and ATAT implement the conventional formalism and provide mature tools for cluster enumeration, model fitting, and Monte Carlo sampling \cite{angqvist_icet_2019, puchala_casm_2023, vandewalle_atat_2002, vandewalle2009,barroso-luque2022}. None of them support a learned chemical embedding, so the number of species they can treat remains limited by the growth in the number of cluster functions.

Here, we introduce \textit{pyeCE}, a general and extensible Python library for graphics-processing-unit (GPU) accelerated Monte Carlo simulations of high-entropy alloys with an arbitrary number of alloying elements. \textit{pyeCE} implements the eCE formalism and provides the full suite of tools needed to parameterize on-lattice cluster expansions. These include the mapping of atomic structures to ideal lattices and the assembly of an eCE model from clusters of interest and symmetry-adapted site descriptors, with flexible chemical embeddings and neural-network architectures. The library also builds training sets, trains the model, and quantifies the uncertainty of its predictions. Built on the \textit{pymatgen}\cite{ong2013} and \textit{PyTorch}\cite{paszke2019pytorch} libraries, \textit{pyeCE} gives users access to a broad materials-science software ecosystem and supports systems with multiple species on multiple sublattices. A trained model can be compiled for fast evaluation on CUDA-enabled GPUs, and finite-temperature properties are obtained from Monte Carlo simulations in several ensembles, with automatic assessment of the confidence interval of each predicted quantity. The modular structure of \textit{pyeCE} separates the description of the lattice, the energy model, and the sampling algorithm, so that the same infrastructure can be reused to build cluster-expansion models of properties beyond bulk chemical thermodynamics.

\section{Building and deploying eCE models with \textit{pyeCE}}
\label{sec:building-ece-models}

\begin{figure}[!htbp]
    \centering
    \includegraphics[width=0.905\linewidth]{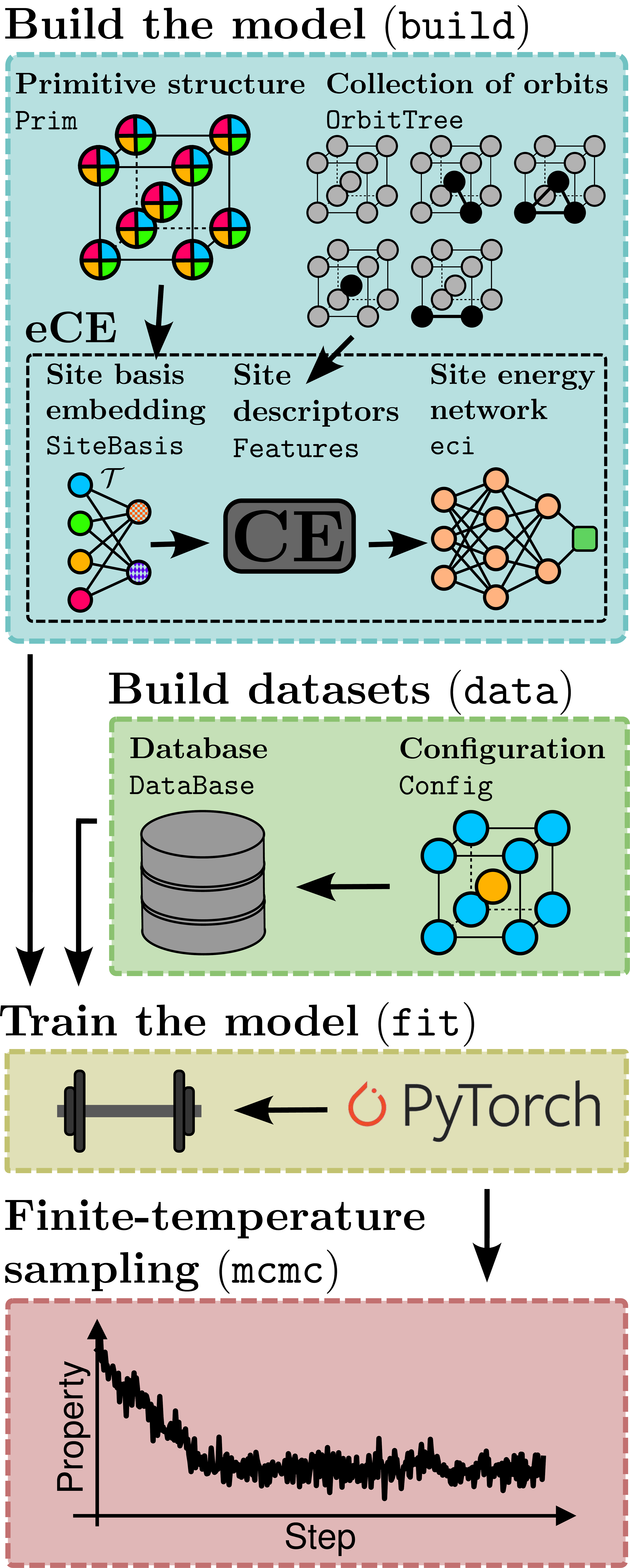}
    \caption{General workflow of \textit{pyeCE}. (1) The eCE model is constructed from the site-basis embedding, the site descriptors, and the site energy model. (2) A training set of symmetrically distinct orderings is enumerated, evaluated with first-principles calculations, and mapped onto the parent lattice. (3) The model is trained using PyTorch by simultaneously optimizing the transformation matrix $\mathcal{T}$ and the site energy model. (4) Finite-temperature properties are predicted by Monte Carlo simulations.}
    \label{fig:workflow_pyece}
\end{figure}

Predicting finite-temperature properties with an embedded cluster expansion proceeds in four stages, namely building the eCE model, assembling a training set, training the model, and sampling thermodynamic properties. \textit{pyeCE} unifies these stages in a single modular workflow, shown in \cref{fig:workflow_pyece}, that mirrors the structure of the established cluster expansion packages described above. The library implements the embedded cluster expansion formalism \cite{muller2025}, which extends on-lattice cluster expansions to alloys with many chemical species, and accelerates model evaluation and Monte Carlo sampling on GPUs. Each stage is accessible through both a command-line interface and a Python API. The remainder of this section describes each stage in detail.

\subsection{Building the eCE Model}

On-lattice cluster expansion (CE) models are parameterized to reproduce how a property depends on the chemical ordering over a \emph{parent} crystal structure. In an eCE model, an extensive property such as the total energy is expressed as a sum of site contributions. We represent any arrangement of elements over the lattice sites as an occupation vector $\vec{\sigma} = (\sigma_1, \sigma_2, \ldots)$, where $\sigma_{i}$ is the index of the chemical species occupying site $i$. The property $x(\vec{\sigma})$ is then:
\begin{equation}
    x(\vec{\sigma}) = \sum_i x_i(\vec{\sigma})
    \label{eq:site_property}
\end{equation}
where $x_i$ is the site property at site $i$. \Cref{eq:site_property} ensures that $x(\vec{\sigma})$ scales with system size.

The occupation vector $\vec{\sigma}$ encodes the chemical arrangement, but it is more convenient to work with descriptors of local ordering built from atomic clusters centered on site $i$. In both the eCE model and conventional cluster expansion models, these descriptors are stratified into contributions from points (one-body), pairs (two-body), triplets (three-body), and higher-order clusters. Clusters related by a space-group symmetry operation of the primitive crystal structure form an \emph{orbit}. A descriptor is made symmetry-invariant by summing the contribution of a cluster over all clusters in its orbit that radiate from site $i$.

\begin{figure}[!ht]
    \centering
    \includegraphics[width=0.99\linewidth]{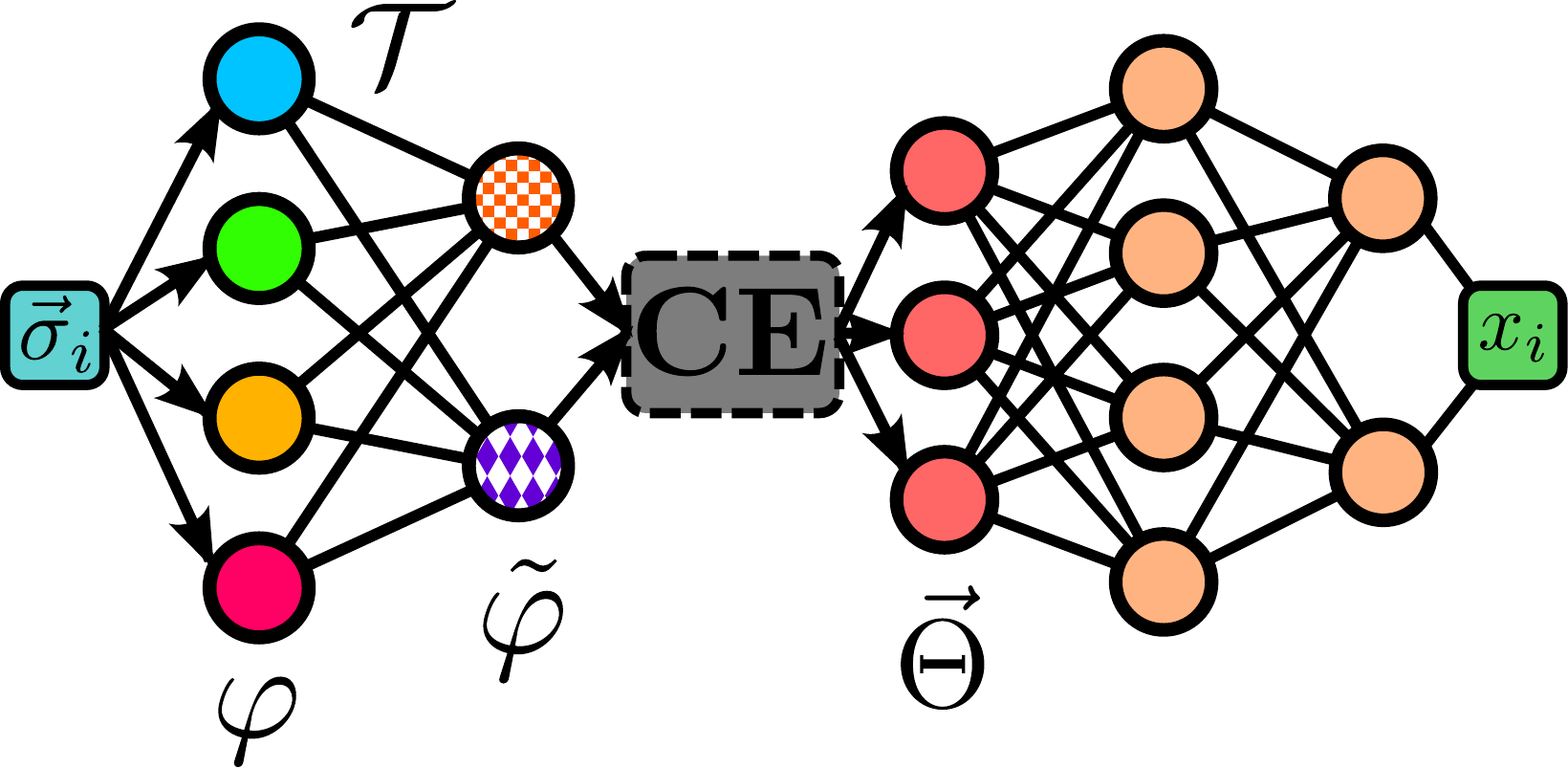}
    \caption{Schematic of the embedded cluster expansion formalism. Given the neighborhood configuration $\vec{\sigma}_i$ of site $i$, the site-basis functions $\varphi$ are projected onto the embedded site-basis functions $\tilde{\varphi}$ through the learned transformation $\mathcal{T}$. Symmetry-invariant site descriptors $\vec{\Theta}$ are then computed as in a conventional cluster expansion. These descriptors serve as input to a feedforward neural network that evaluates the site property $x_i$. The dependence of these quantities on the configuration is suppressed for clarity.}
    \label{fig:eCE_energy_model}
  \end{figure}

  \Cref{fig:eCE_energy_model} illustrates the essential steps taken within \textit{pyeCE} to predict $x_i$ at each site $i$. First, the site-basis functions are evaluated for the species occupying each site in the neighborhood of site $i$. These are then projected into a lower-dimensional space. Tensor products of the projected site-basis functions are symmetrized by summing over all cluster functions related by symmetry, yielding symmetry-invariant descriptors. These descriptors serve as inputs to a neural network that computes $x_i$.

  \textit{pyeCE} provides a set of modular classes for constructing the elements of the eCE model illustrated in \cref{fig:eCE_energy_model}. The \texttt{Prim} object encodes the primitive crystal structure and the chemical degrees of freedom at each site, as shown in \cref{fig:workflow_pyece}. Clusters are enumerated up to a user-defined maximum size (number of atoms including the central site) and a maximum pairwise distance between sites. The \texttt{OrbitTree} object holds all clusters included in the eCE model. The \texttt{build} command automates the construction of the embedding, the site descriptors, and the computational graph of the neural-network energy model.

\subsubsection{Embedding the Chemical Species}

As in a conventional cluster expansion, the state of a site in a $c$-component alloy is described by $c$ linearly independent site-basis functions $\varphi_0, \varphi_1, \ldots, \varphi_{c-1}$. Their values evaluated on the $c$ species form the $c \times c$ chemical embedding matrix $\bm{\varphi}$, with elements $\bm{\varphi}_{ij} = \varphi_i(j)$, where $\varphi_i$ is the $i$th site-basis function and $j$ indexes the species. Each row of $\bm{\varphi}$ lists the values of one site-basis function across the $c$ species, and each column gives the values of all $c$ site-basis functions for a single species.

The eCE formalism then embeds the $c$ site-basis functions in a lower-dimensional space via a linear transformation:
\begin{equation}
    (\tilde{\varphi}_0,\, \tilde{\varphi}_1, \ldots ,\, \tilde{\varphi}_{k-1})^T = \mathcal{T} (\varphi_0,\, \varphi_1,\, \ldots ,\, \varphi_{c-1})^T
    \label{eq:transformation}
\end{equation}
The transformation matrix $\mathcal{T} \in \mathbb{R}^{k \times c}$ projects the $c$ site-basis functions onto $k < c$ \textit{embedded} functions $\tilde{\varphi}$. The embedded site-basis functions no longer span the full compositional space, so the $c$ vectors mapping the $c$ chemical species onto $\mathbb{R}^k$ become linearly dependent. The transformed embedding matrix is $\tilde{\bm{\varphi}} = \mathcal{T} \bm{\varphi} \in \mathbb{R}^{k \times c}$, and the $c$-component system is effectively treated as a pseudo-$k$-component system.

\begin{figure}[!ht]
    \centering
    \includegraphics[width=0.99\linewidth]{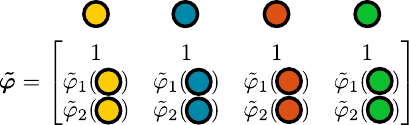}
    \caption{Illustration of the embedded site-basis functions that map $c = 4$ species onto $k = 3$. The four chemical species are represented by four colors (yellow, blue, red, and green). Each species maps to one of four linearly dependent vectors $(1,\, \tilde{\varphi}_1,\, \tilde{\varphi}_2)^T$ in $\mathbb{R}^3$.}
    \label{fig:embedding}
\end{figure}

\Cref{fig:embedding} shows an example with $c = 4$ species mapped to $k = 3$ embedded site-basis functions. The zeroth function is always set to the constant $\tilde{\varphi}_0 = 1$, following the conventional CE formalism. This choice enables a hierarchical expansion organized by the number of atoms in each cluster. 

The transformation matrix $\mathcal{T}$ is learned during model training to identify the embedded site-basis functions that best capture the chemical similarities among the elements in the training set. Whereas conventional CE models treat $c$ species as $c$ linearly independent vectors, fitting $\mathcal{T}$ encodes the dependencies among species and effectively reduces the number of independent chemical degrees of freedom.

Throughout the remainder of this section, we use the embedded site-basis functions $\tilde{\varphi}_\nu$ to describe the eCE framework and its implementation in \textit{pyeCE}. A conventional cluster expansion follows as a special case by setting $\mathcal{T}$ to the identity matrix, and therefore entails no loss of generality.

\subsubsection{Constructing Site Descriptors}

Local ordering descriptors are constructed by taking products of site-basis functions across multiple sites. \emph{Cluster functions}, denoted $\tilde{\Phi}_{\vec{\alpha}}(\vec{\sigma})$, are computed as:
\begin{equation}
    \tilde{\Phi}_{\vec{\alpha}} (\vec{\sigma}) = \prod_{(i,\nu) \in \vec{\alpha}} \tilde{\varphi}_\nu (\sigma_i)
    \label{eq:cluster_functions}
\end{equation}
where $\vec{\alpha}$ is a list of $N$ tuples, each pairing a site index $i$ in the cluster with a site-basis function index $\nu$. When $c$ linearly independent site-basis functions are used to construct the cluster functions of \cref{eq:cluster_functions}, they form a complete basis\cite{sanchez1984}, and any function of the configuration $\vec{\sigma}$ can be expanded in terms of this basis. A scalar property $x$ invariant under the space-group symmetry operations of the parent crystal structure can then be written as:
\begin{equation}
    x(\vec{\sigma}) = \sum_{\Omega_{\vec{\alpha}}} J_{\Omega_{\vec{\alpha}}} \sum_{\vec{\beta} \in {\Omega_{\vec{\alpha}}}} \tilde{\Phi}_{\vec{\beta}} (\vec{\sigma}) = \sum_{\Omega_{\vec{\alpha}}} J_{\Omega_{\vec{\alpha}}} \Theta_{\Omega_{\vec{\alpha}}} (\vec{\sigma})
    \label{eq:cluster_expansion}
  \end{equation}
where ${\Omega_{\vec{\alpha}}}$ is the orbit of cluster functions related by symmetry and $\Theta_{\Omega_{\vec{\alpha}}}$ are the corresponding symmetry-invariant descriptors. The coefficients $J_{\Omega_{\vec{\alpha}}}$, called \textit{effective cluster interactions} (ECI), are learned from a training set of first-principles calculations. When interactions are short-ranged, the expansion can be truncated to clusters whose atoms lie within a cutoff radius.

Site-centric local descriptors of ordering can be constructed in a manner similar to $\Theta_{\Omega_{\vec{\alpha}}}$ to parameterize the site contributions $x_i$ in \cref{eq:site_property}\cite{natarajan2018}:
\begin{equation}
    \Theta_{\Omega_{\vec{\alpha}}^i} (\vec{\sigma}) = \sum_{\vec{\beta} \in {\Omega_{\vec{\alpha}}^i}} \prod_{(j,\nu) \in \vec{\beta}} \tilde{\varphi}_\nu (\sigma_j)
    \label{eq:site_descriptors}
\end{equation}
where $\Omega_{\vec{\alpha}}^i$ is the set of symmetrically equivalent cluster functions built from clusters that radiate from site $i$.

\begin{figure}[!ht]
    \centering
    \includegraphics[width=0.99\linewidth]{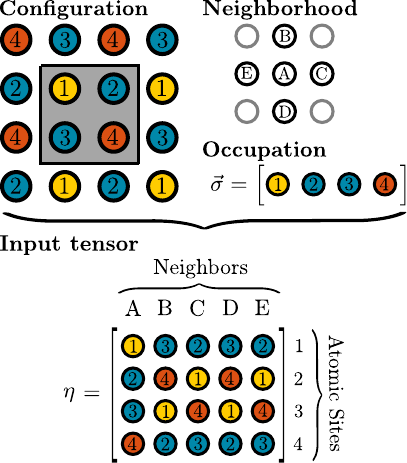}
    \caption{Illustration of the construction of the input matrix $\eta$. The example configuration contains 4 atomic sites (1, 2, 3, and 4). The neighborhood of each site consists of the site itself (A) and its 4 nearest neighbors (B, C, D, and E). Element $\eta_{ij}$ of the input matrix holds the chemical species at neighborhood position $j$ of the $i$th atomic site.}
    \label{fig:input_tensor}
\end{figure}

\textit{pyeCE} implements cluster expansions based on the site-centric descriptors of \cref{eq:site_descriptors}. The neighborhood of site $i$ consists of all sites that form a cluster with it for which a cluster function contributes to the local ordering descriptor. The chemical arrangement within the neighborhood of each site is collected in a matrix $\eta$, where each row corresponds to one central site and lists the occupancy of that site followed by the occupancies of its neighbors in a fixed order. In the example of \cref{fig:input_tensor}, 4 central sites (labeled 1--4) each have a neighborhood of 5 sites (A--E). The matrix $\eta$ is therefore $4 \times 5$, with element $\eta_{ij}$ giving the occupancy of the $j$th neighbor of the $i$th central site. This matrix contains all the information needed to evaluate the site property $x_i$ across the crystal.

\begin{figure}[!ht]
    \centering
    \includegraphics[width=0.99\linewidth]{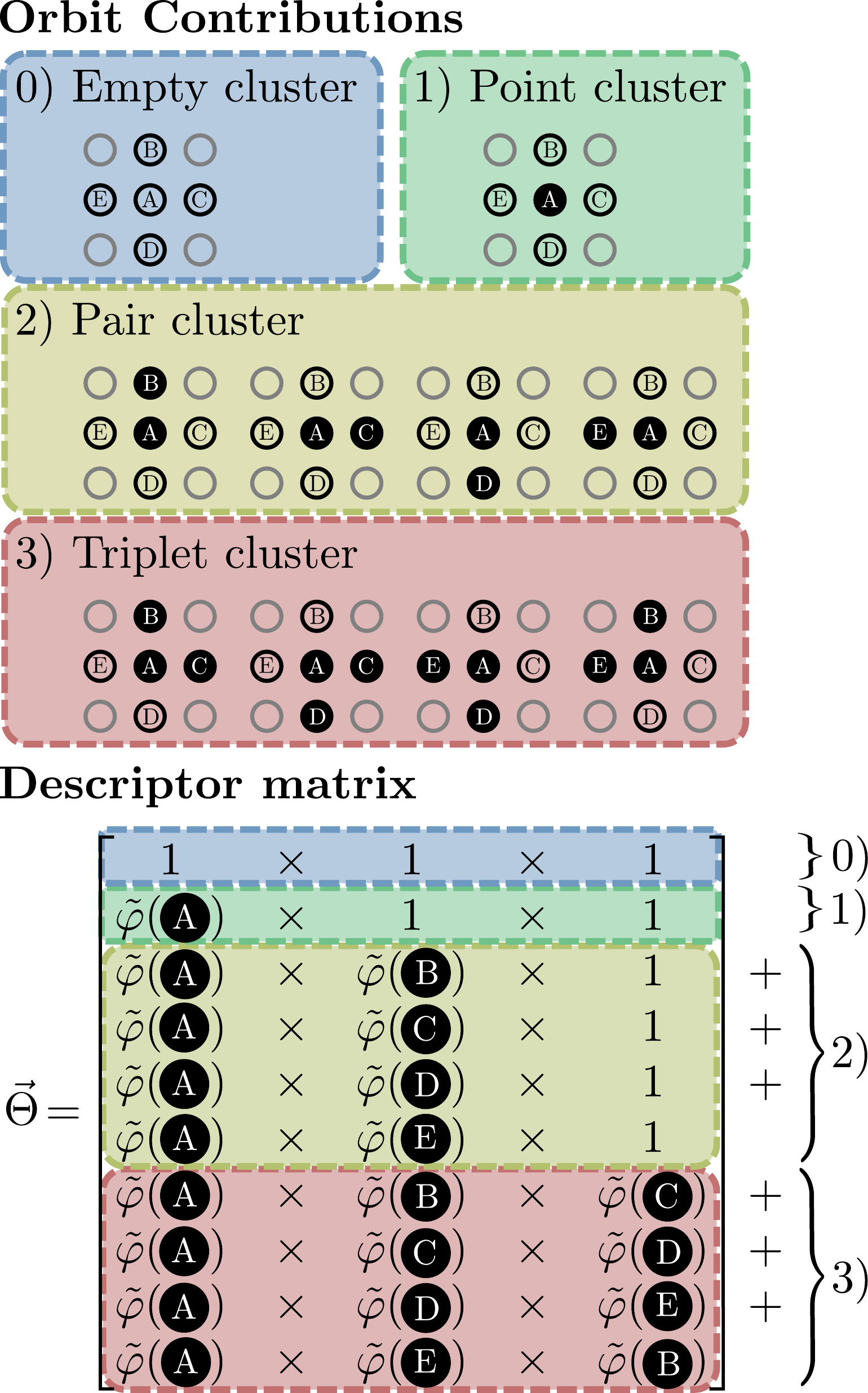}
    \caption{Illustration of the construction of symmetry-invariant site descriptors for an alloy with two site-basis functions on a square lattice. The toy eCE model includes four orbits: the empty, point, nearest-neighbor pair, and smallest triplet clusters. The neighborhood consists of the central site (A) and its nearest neighbors (B, C, D, and E). The descriptor matrix collects the cluster functions of all clusters in the site-centric orbits, with shorter clusters padded by the constant 1. Site descriptors $\vec{\Theta}$ are computed by taking the product of the entries in each row and summing rows belonging to the same orbit.}
    \label{fig:descriptors}
\end{figure}

\textit{pyeCE} builds the site descriptors efficiently through tensor operations in PyTorch. For each central site, the site-basis function values for all relevant clusters are collected in a descriptor matrix, where each row corresponds to one cluster function $\tilde{\Phi}_{\vec{\alpha}}$ and each column to a site in the cluster. The number of columns equals the body-order of the largest cluster, and shorter clusters are padded with the constant $\tilde{\varphi}_0 = 1$. Because site-centric neighborhoods share a fixed ordering, this matrix is constructed once at model initialization using a lookup table.

Cluster functions are obtained by taking the product across all entries in each row of the descriptor matrix. Symmetrized descriptors are then computed by summing all rows that belong to the same orbit. Grouping clusters of the same orbit consecutively allows this summation to be performed via scatter-reduction operations, enabling efficient GPU acceleration.

\Cref{fig:descriptors} illustrates this construction for an alloy with two site-basis functions on a square lattice with empty, point, nearest-neighbor pair, and triplet clusters. The four orbits yield four entries in the site descriptor vector $\vec{\Theta}$.

The site descriptor construction contains no learnable parameters, but depends on two classes of hyperparameters, namely the number of \textit{embedded} site-basis functions $k$ and the set of clusters to include. Together, the chemical embedding and the site descriptor construction act as an encoder that maps the local chemical arrangement into a fixed-dimensional, symmetry-invariant representation of the local chemical ordering.

\subsubsection{Evaluating the Site Property}

\begin{figure}[!ht]
    \centering
    \includegraphics[width=0.8\linewidth]{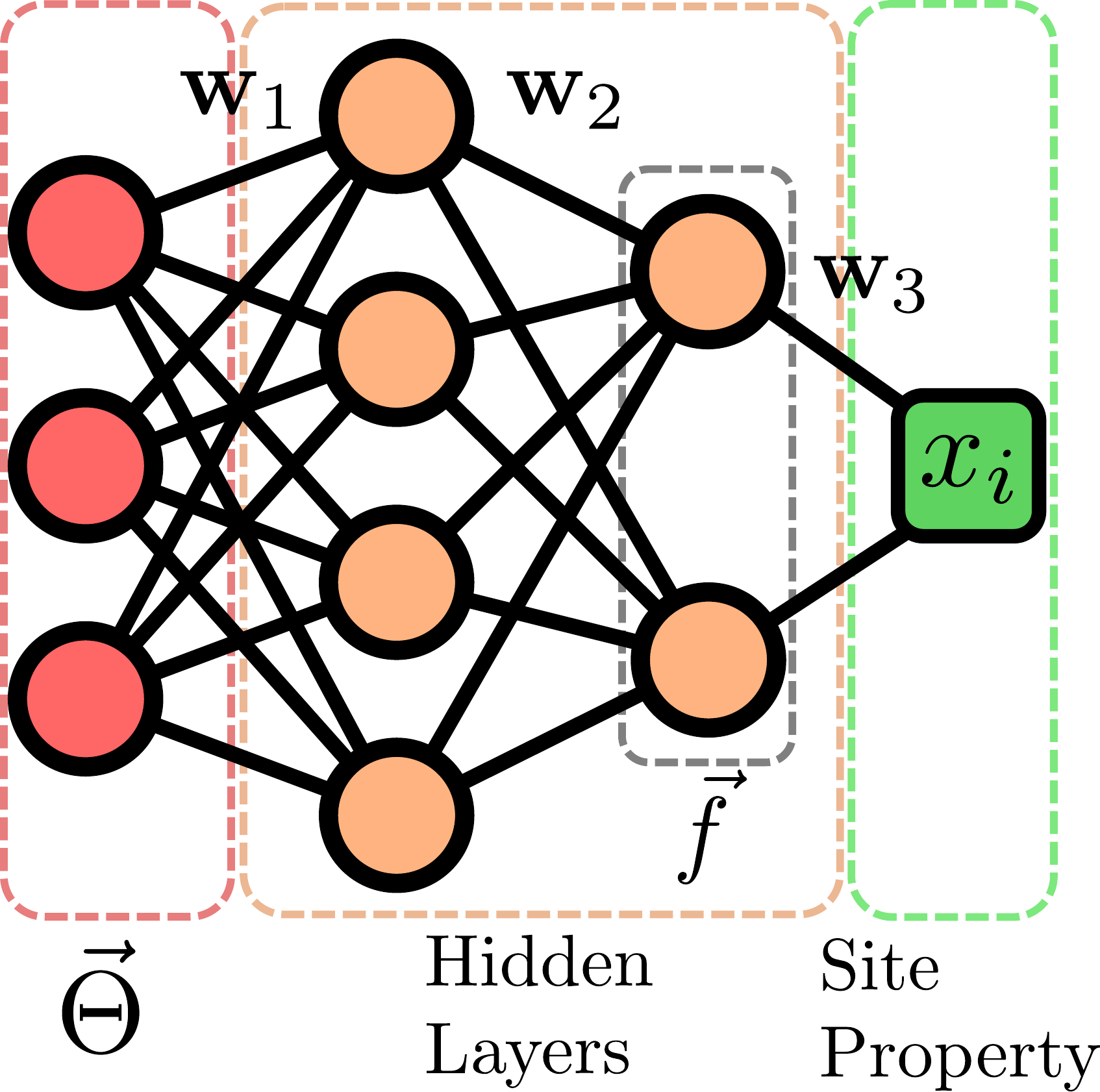}
    \caption{Illustration of a site property feedforward neural network. The input layer corresponds to the site descriptors $\vec{\Theta}$. The neural network is composed of 2 hidden layers. The final hidden layer produces the latent features $\vec{f}$. The final neuron outputs the site property $x_i$. The weights $\textbf{w}=\{ \textbf{w}_1, \textbf{w}_2, \textbf{w}_3 \}$ and biases in each layer are the trainable parameters.}
    \label{fig:site_model}
\end{figure}

A feedforward neural network takes the site descriptors $\vec{\Theta}$ as input and evaluates the site property $x_i$ at site $i$, as illustrated in \cref{fig:site_model}. The total property $x$ is then the sum of all site contributions, in accordance with \cref{eq:site_property}. This approach generalizes the conventional linear CE model of \cref{eq:cluster_expansion}, which is equivalent to a single-layer network whose weights are the ECI coefficients $J_{\Omega_{\vec{\alpha}}}$, and follows the framework of \citeauthor{natarajan2018}~\cite{natarajan2018}.

\subsection{Assembling a Training Set}

Parameterizing an eCE model requires a training set of symmetrically distinct chemical arrangements of the $c$ species over the parent crystal structure and their associated energies. Symmetrically distinct arrangements can be systematically enumerated with established algorithms\cite{hart_enumeration_2008}, and their formation energies or other properties of interest evaluated by first-principles calculations, typically based on density functional theory (DFT). These structure--energy pairs constitute the training set.

Properties of interest, such as formation energies, are typically evaluated for the relaxed chemical arrangement, in which ionic and cell relaxation lower the total energy to a local minimum, so the lattice vectors and atomic coordinates differ from those of the ideal structure. On-lattice models assume atoms sit on ideal sites, so each relaxed structure must be mapped onto the parent lattice. Structures for which this mapping is ambiguous or poorly conditioned must be identified and discarded before training.

Mapping a child structure onto a parent structure is a two-step process that first matches lattices and then assigns atomic sites \cite{thomas_comparing_2021}. The parent lattice $L$ and child lattice $S$ are related by an integer transformation matrix $T$ and a deformation matrix $D$ such that:
\begin{equation}
        L \cdot T = D \cdot S
        \label{eq:lattice_mapping}
\end{equation}
Here, $L=[\vec{l}_1,\, \vec{l}_2,\, \vec{l}_3]$ and $S=[\vec{s}_1,\, \vec{s}_2,\, \vec{s}_3]$ are $3 \times 3$ matrices containing the parent and child lattice vectors, respectively, and $\det(T)$ equals the ratio of the number of atoms in the child cell to the number in the parent primitive cell.

The robust crystal structure mapping algorithm of \citeauthor{thomas_comparing_2021} \cite{thomas_comparing_2021} rigorously maps a deformed child structure onto a reference parent crystal structure. However, it can be computationally slow, particularly for the small-distortion structures that make up a large fraction of training sets. \textit{pyeCE} provides a faster structure matching algorithm that is valid when the child and parent structures share nearly the same orientation and lattice parameters, so that $D \approx I$. In this limit, \cref{eq:lattice_mapping} becomes:
\begin{equation*}
    L \cdot T = (I + dD) \cdot S
\end{equation*}
where $dD = D - I$ is the deviation of $D$ from the identity. Left-multiplying by $L^{-1}$ and rearranging gives:
\begin{equation*}
    L^{-1} \cdot S = T - L^{-1} \cdot dD \cdot S
\end{equation*}
Since $T$ is an integer matrix by definition, $L^{-1} \cdot S \approx T$ up to the small perturbation $L^{-1} \cdot dD \cdot S$. Finding $T$ is therefore a closest vector problem (CVP), namely identifying the lattice point in $L$ nearest to each child lattice vector,
\begin{equation}
    T_{i1}, T_{i2}, T_{i3} = \argmin_{T_{ij} \in \mathbb{Z}} \left\lVert \sum_{j=1}^3 T_{ij} \vec{l}_j - \vec{s}_i \right\rVert \, , \; \forall \; i=1,2,3
\end{equation}
This problem is NP-hard. Babai's rounding algorithm finds an approximate solution by rounding each element of $L^{-1} \cdot S$ to the nearest integer \cite{babai_lovasz_1986}, and performs better when the parent lattice vectors form a Lenstra--Lenstra--Lov\'asz (LLL)-reduced basis \cite{lenstra_factoring_1982}.

\begin{figure}[!ht]
    \centering
    \includegraphics[width=0.99\linewidth]{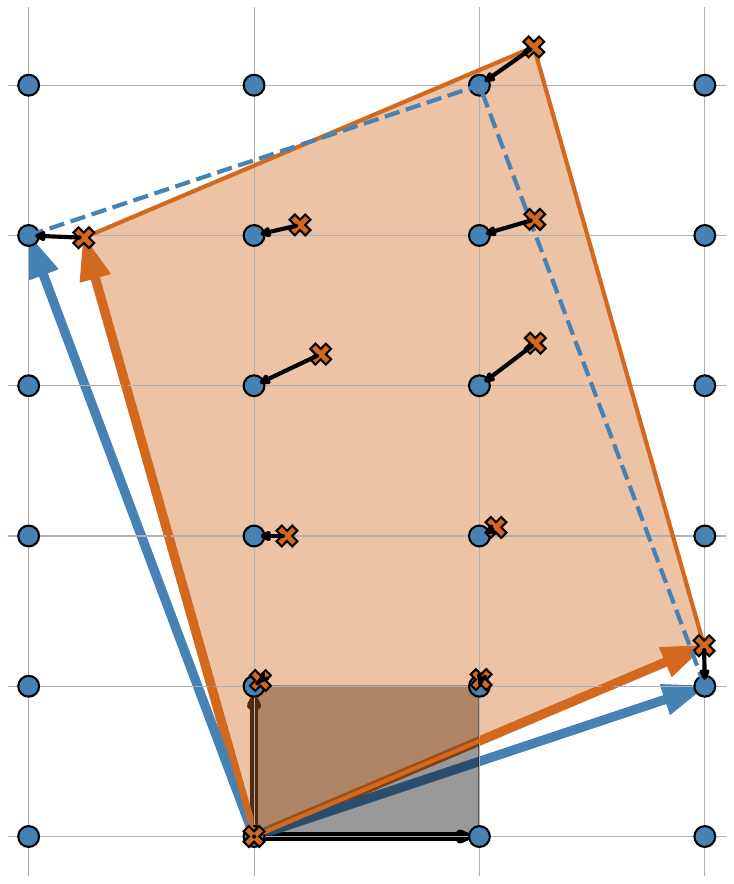}
    \caption{Illustration of the fast structure matching algorithm. The gray cell is the parent structure and the blue dots are the parent lattice sites. The orange cell is the child structure, whose lattice vectors are shown as orange arrows. The mapping rounds each child lattice vector to the nearest parent lattice point, then assigns each child atomic site to the nearest parent site.}
    \label{fig:mapping}
\end{figure}

\Cref{fig:mapping} illustrates the mapping process. First, the lattice vectors of the child structure are matched to the nearest parent lattice points by rounding $L^{-1} \cdot S$ to the nearest integer.

Misorientation embeds a rotation in $D$ whose off-diagonal elements are not small, and the algorithm can then fail.

Once the lattice is matched, atomic sites are assigned by solving a linear assignment problem following the methodology of \citeauthor{thomas_comparing_2021}, which minimizes the total displacement between child and parent sites. Each mapped structure is then assigned a lattice deformation score and an atomic displacement score using the same metrics. Structures whose scores exceed user-defined thresholds are discarded from the training set. Because \textit{pyeCE} uses the approximate CVP solution rather than full enumeration, these scores are conservative. They may overestimate the true mapping cost, but never underestimate it, so no poorly mapped structure is incorrectly accepted. The fast algorithm can therefore be applied first, with structures of high deformation score passed to the full algorithm in a second pass to recover any configurations that the approximation may have rejected.

\subsection{Training the Model}

Having built the eCE model and assembled the training set, we optimize the model parameters to reproduce the training data. The neural network and the learnable \textit{embedded} site-basis functions render the model nonlinear, precluding the linear least-squares fitting used in conventional cluster expansions. The model is trained by minimizing the loss function
\begin{equation}
    \mathcal{L}(\mathcal{T}, \textbf{w}) =\sum_i \big(x(\vec{\sigma}_i) - x_{eCE}(\vec{\sigma}_i ; \mathcal{T}, \textbf{w}) \big)^2 + \mathcal{L}_{reg}(\mathcal{T}, \textbf{w})
    \label{eq:model_optimization}
\end{equation}
where \textbf{w} are the weights of the neural-network site property model and $\mathcal{L}_{reg}$ is a regularization term. Training uses gradient-descent optimizers from PyTorch. Overfitting is controlled by $L_2$ regularization (weight decay), early stopping, and learning-rate scheduling. In addition to these standard tools, the training proceeds iteratively, progressively incorporating descriptors for larger clusters. This \textit{ladder training} improves robustness and helps avoid local minima.

In practice, eCE models containing only pair clusters with cutoff radii between 8\,\AA\ and 10\,\AA\ and the first few triplet clusters may be sufficient. Ladder training is typically divided into three to six steps, with the first step containing only the first and second nearest-neighbor pair clusters. This step accounts for the largest reduction in loss and can use a relatively large learning rate, around $10^{-2}$.

\subsubsection{Estimating Prediction Errors}

Prediction errors are inevitable in any surrogate model, and reliable error estimates are important for assessing the confidence of individual predictions. \textit{pyeCE} implements two techniques to estimate the prediction uncertainty of a trained model.

The first is the committee method, in which multiple models are trained with identical hyperparameters but different random initial weights. Each model converges to a slightly different solution because of the initialization and the stochasticity of gradient descent. An error estimate $\hat{\sigma}_{\epsilon}^{CM}(\vec{\sigma})$ for configuration $\vec{\sigma}$ is obtained from the root-mean-square of their deviations from the committee mean:
\begin{equation}
    \hat{\sigma}_{\epsilon}^{CM}(\vec{\sigma}) = \sqrt{\frac{1}{N}\sum_{i=1}^N \epsilon_i^2(\vec{\sigma})}
\end{equation}
where $\epsilon_i(\vec{\sigma}) = x_{eCE}^i(\vec{\sigma}) - \bar{x}_{eCE}(\vec{\sigma})$ is the deviation of the $i$th model from the committee average $\bar{x}_{eCE}(\vec{\sigma}) = \frac{1}{N}\sum_{i=1}^N x_{eCE}^i(\vec{\sigma})$. Although robust and straightforward, this method is computationally demanding, as it requires training multiple models and evaluating each configuration with all of them.

The second method, following \citeauthor{bigi_prediction_2024} \cite{bigi_prediction_2024}, estimates the error from the latent features $\vec{f}$ of the last layer of the trained network (\cref{fig:site_model}). The error estimate $\hat{\sigma}_{\epsilon}^{LL}(\vec{\sigma})$ is given by:
\begin{equation}
   \hat{\sigma}_{\epsilon}^{LL}(\vec{\sigma}) = \sqrt{\alpha \vec{f}^T(\vec{\sigma}) (F^T F + \xi^2 I)^{-1} \vec{f}(\vec{\sigma})}
\end{equation}
where $F$ is a matrix whose $i$th row holds the last-layer features of the network evaluated at the $i$th training configuration, $\alpha$ is a calibration parameter set so that the mean error estimate matches the training root-mean-square error (RMSE), and $\xi$ is a small regularization parameter. This method requires only a single training run and minimal additional computation.

\subsection{Sampling Thermodynamic Properties}

eCE models are sampled with the same statistical-mechanics techniques as conventional cluster expansions, giving access to thermodynamic properties, free energies, and phase stability as functions of temperature and composition. For a thermodynamic ensemble with energy $\omega$, the probability of observing configuration $\vec{\sigma}$ is:
\begin{equation}
\begin{split}
    \mathbb{P}(\vec{\sigma}) &= \frac{1}{Z} \exp\big(-\beta \omega(\vec{\sigma}) \big) \\
    \text{with} \quad Z &= \sum \exp\big(-\beta \omega(\vec{\sigma}) \big)
\end{split}
\label{eq:ensemble_probability}
\end{equation}
where the sum runs over all microstates commensurate with the boundary conditions of the ensemble.

\begin{figure}[!ht]
    \centering
    \includegraphics[width=0.99\linewidth]{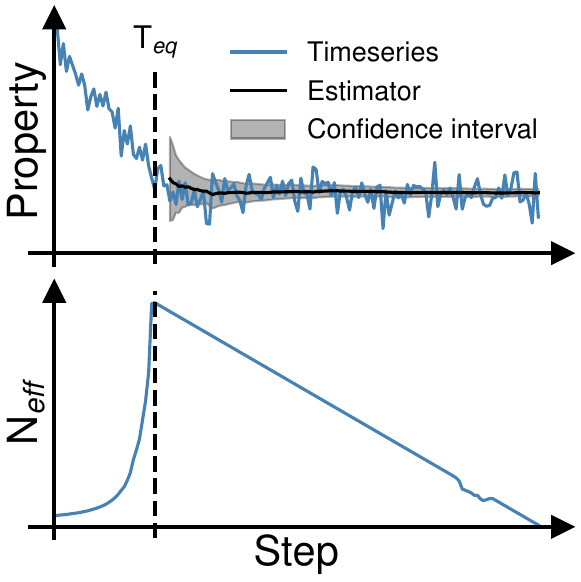}
    \caption{Illustration of a time series resulting from a Metropolis Monte Carlo simulation. The upper plot shows the evolution of the property with the number of MC steps performed, together with the sample-mean estimator $\hat{\mu}_X$ and its confidence interval. The lower plot shows the number of effective uncorrelated samples $N_{eff}$ contained in the time series from a given MC step to the end. The equilibration time corresponds to the MC step with the largest number of effective samples.}
    \label{fig:timeseries}
\end{figure}

Markov chain Monte Carlo (MCMC) methods such as the Metropolis algorithm \cite{metropolis1953, hastings_1970} construct an ergodic Markov chain whose configurations sample $\mathbb{P}$, so that ensemble averages become time averages along the chain:
\begin{equation}
    \langle X \rangle = \sum_{\vec{\sigma}} X(\vec{\sigma}) \mathbb{P}(\vec{\sigma}) \approx \underset{N\to \infty}{\lim}\frac{1}{N}\sum_i^N X_{i}
    \label{eq:mc_average}
\end{equation}
where $X_i = X(\vec{\sigma}_i)$ and $\vec{\sigma}_i$ is the $i$th configuration drawn from $\mathbb{P}$ (\cref{eq:ensemble_probability}). Each step proposes a new configuration according to a user-defined proposal distribution, typically swapping two atoms in the canonical ensemble or changing the species on one site in the semi-grand canonical ensemble. Evaluating the property $X_t$ at every step $t$ yields a time series, such as the one in \cref{fig:timeseries}.

The time average in \cref{eq:mc_average} is the sample-mean estimator $\hat{\mu}_X$ of the ensemble average $\langle X \rangle$. It is unbiased and converges to $\langle X \rangle$ as the number of samples grows. Two features of MCMC data complicate its use. First, successive configurations are correlated. Second, the chain begins away from equilibrium, so its early samples must be discarded. \textit{pyeCE} accounts for both before reporting an average and its confidence interval.

As the chain produces correlated samples, the variance of $\hat{\mu}_X$ is not simply $\sigma_X^2/N$. For a time series it is:
\begin{equation}
    \begin{split}
        \mathbb{V}[\hat{\mu}_{X}] &= \frac{1}{N} \sigma_{X}^2 \Big(1 + 2 \sum_{n=1}^{N-1} \big(1-\frac{n}{N}\big) C_n\Big) \\
        \text{with}\quad C_n &= \frac{\text{Cov}(X_i, X_{i+n})}{\sigma_{X}^2}
    \end{split}
    \label{eq:variance_mean_estimator}
\end{equation}
where $\sigma_{X}^2$ is the variance of the property $X$ and $C_n$ is the autocorrelation at lag $n$. The first term is the variance of the sample mean for independent samples. The second term corrects for correlation between samples. When the autocorrelation is positive, this correction is positive and $\mathbb{V}[\hat{\mu}_{X}]$ decays more slowly than the independent-sample rate of $1/N$ \cite{walle2002, chodera_equilibration_time_2016}. Equivalently, the time series carries only $N_{eff} = \sigma_{X}^2 / \mathbb{V}[\hat{\mu}_{X}]$ effective uncorrelated samples \cite{chodera_equilibration_time_2016}. Correlation can be handled either by evaluating \cref{eq:variance_mean_estimator} directly or by retaining every $N/N_{eff}$th sample and applying the standard estimators for uncorrelated samples.

Averages must be computed only over the equilibrated portion of the chain. Samples drawn before the chain reaches its equilibrium distribution reflect the initial configuration and bias the average. \textit{pyeCE} locates the equilibration time with the algorithm of \citeauthor{chodera_equilibration_time_2016} \cite{chodera_equilibration_time_2016}. For each step $t = 1, 2, \ldots, N$, the algorithm computes $N_{eff}$ over the window from $t$ to $N$ and takes the equilibration time as the step that maximizes $N_{eff}$. \Cref{fig:timeseries} illustrates this construction.

From the equilibrated, decorrelated samples, \textit{pyeCE} reports $\hat{\mu}_X$ together with a confidence interval that quantifies convergence. The interval is constructed at a user-specified confidence level from the variance in \cref{eq:variance_mean_estimator}, assuming that $\hat{\mu}_{X}$ follows a $t$-distribution. \Cref{fig:timeseries} schematically shows this estimator and its confidence interval as a function of MC step.

\textit{pyeCE} evaluates ensemble averages of configuration-dependent properties on the fly. At each uncorrelated sample it evaluates all requested properties and continues the simulation until they converge within a user-specified tolerance. Properties that are already implemented include sublattice compositions for long-range order (LRO) parameters, cluster occupation probabilities, and the associated short-range order (SRO) parameters.

\section{Applications}

\textit{pyeCE} is a modular codebase for studying the finite-temperature thermodynamics of complex concentrated alloys. We illustrate its capabilities through two case studies. The first characterizes the thermodynamics and short-range order of high-entropy refractory alloys for high-temperature and extreme-environment use. The second predicts hydrogen dissolution in a ternary alloy of current interest for fusion and energy storage.

\subsection{Thermodynamics of a 9-component refractory alloy }

Multicomponent refractory alloys comprising elements in groups 4, 5, and 6 of the periodic table are attractive for their high-temperature properties and tunable mechanical strength. However, questions remain about their stability as a disordered solid solution at lower temperatures, their corrosion resistance, and the nature of short-range order in these systems. The first application studies the thermodynamic properties of BCC alloys in the Cr--Hf--Mo--Nb--Ta--Ti--V--W--Zr system. We parameterized the eCE model by embedding the 9 independent site-basis functions within a three-dimensional space, effectively treating the alloy as a pseudo-ternary system. The first 8 nearest-neighbor pairs and the smallest triplet cluster were considered. The site energy model was a feedforward neural network with 3 hidden layers of 128, 128, and 32 nodes.

\begin{figure}[!ht]
    \centering
    \includegraphics[width=0.99\linewidth]{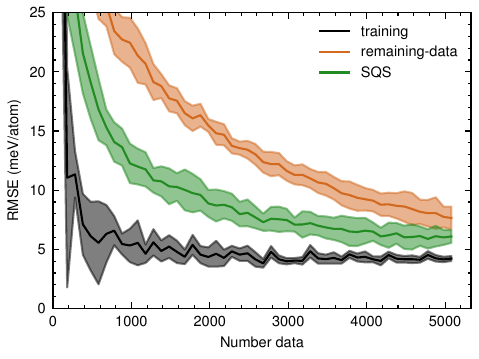}
    \caption{Root-mean-square error (RMSE) of the eCE model for the Cr--Hf--Mo--Nb--Ta--Ti--V--W--Zr system as a function of the number of structures in the training set during active learning. Errors are shown for the training set (black), the structures remaining in the data reservoir (orange), and the held-out special quasi-random structure (SQS) test set (green). Solid lines and shaded bands denote the mean and standard deviation of the RMSE over 10 independent training runs.
    }
    \label{fig:9comp_training}
  \end{figure}

  We used the error-estimation tools in \textit{pyeCE} to build a training set spanning the full 9-component composition space by active learning, iteratively adding the structures with the largest predicted errors from a pool of reservoir structures. The initial training set contained the pure elements and all symmetrically distinct decorations of a 2-atom supercell. We constructed a reservoir of candidate structures with the \citeauthor{hart_enumeration_2008} enumeration algorithm \cite{hart_enumeration_2008}. At each iteration, we trained a model on the current training set and evaluated the last-layer uncertainty $\hat{\sigma}_{\epsilon}^{LL}$ of every reservoir structure. We then added the 100 highest-uncertainty structures, together with their DFT energies, to the training set, and repeated until the training set contained 5000 structures.

  To trace how the error evolves during this procedure, we used a large database of DFT calculations from previous studies\cite{muller2025, lee2026}. Because every structure in this database already has a DFT energy, we could track not only the training error but also the prediction error on the reservoir structures not yet added to the training set. eCE models are used primarily to probe disordered solid solutions, so we also assembled a held-out test set of equiatomic special quasi-random structures (SQS). These SQS were never seen during training and therefore measured the accuracy of the model in the disordered state.

\Cref{fig:9comp_training} shows how the training, reservoir, and test errors evolve with training-set size. We performed 10 independent training runs and estimated the mean and standard deviation of the RMSE for each dataset using the committee method described above. The training error converged to $\approx 5$\,meV/atom. The reservoir error converged more slowly, reaching 10\,meV/atom only after about 4000 structures had been added to the training set. The test error on the SQS converged faster than the reservoir error and stabilized near 6\,meV/atom.

\begin{figure}[!ht]
    \centering
    \includegraphics[width=0.99\linewidth]{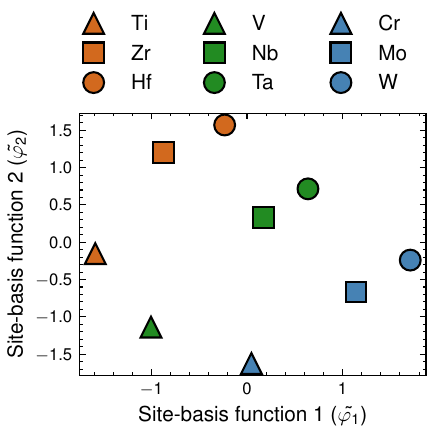}
    \caption{Projection of the 9 chemical species onto the two non-trivial embedded site-basis functions. The color of each marker denotes the periodic-table group of the element and the marker shape denotes its period.}
    \label{fig:9comp_phi}
\end{figure}

The learned transformation matrix $\mathcal{T}$ maps the 9 chemical species into a three-dimensional space with one dimension fixed to the constant $\tilde{\varphi}_0 = 1$. The resulting vectors offer a data-driven measure of the chemical similarities and dissimilarities among the elements. \Cref{fig:9comp_phi} shows the projection of each element onto the two non-trivial embedded site-basis functions. The elements are ordered by group and period, mirroring their arrangement in the periodic table. The $3d$ elements occupy the lower-left region of the space, while the $4d$ and $5d$ elements appear toward the upper right. Elements within each group are also separated. These results indicate that the eCE model recovers chemical relationships consistent with the periodic table.

Having parameterized an accurate model, we performed canonical Monte Carlo simulations for all 502 equiatomic alloys formed from 2 to 9 of the elements in this system, using the same Monte Carlo parameters as in \cite{lee2026}. We used these simulations to characterize the SRO in the disordered state at multiple temperatures.

\begin{figure}[!ht]
    \centering
    \includegraphics[width=0.99\linewidth]{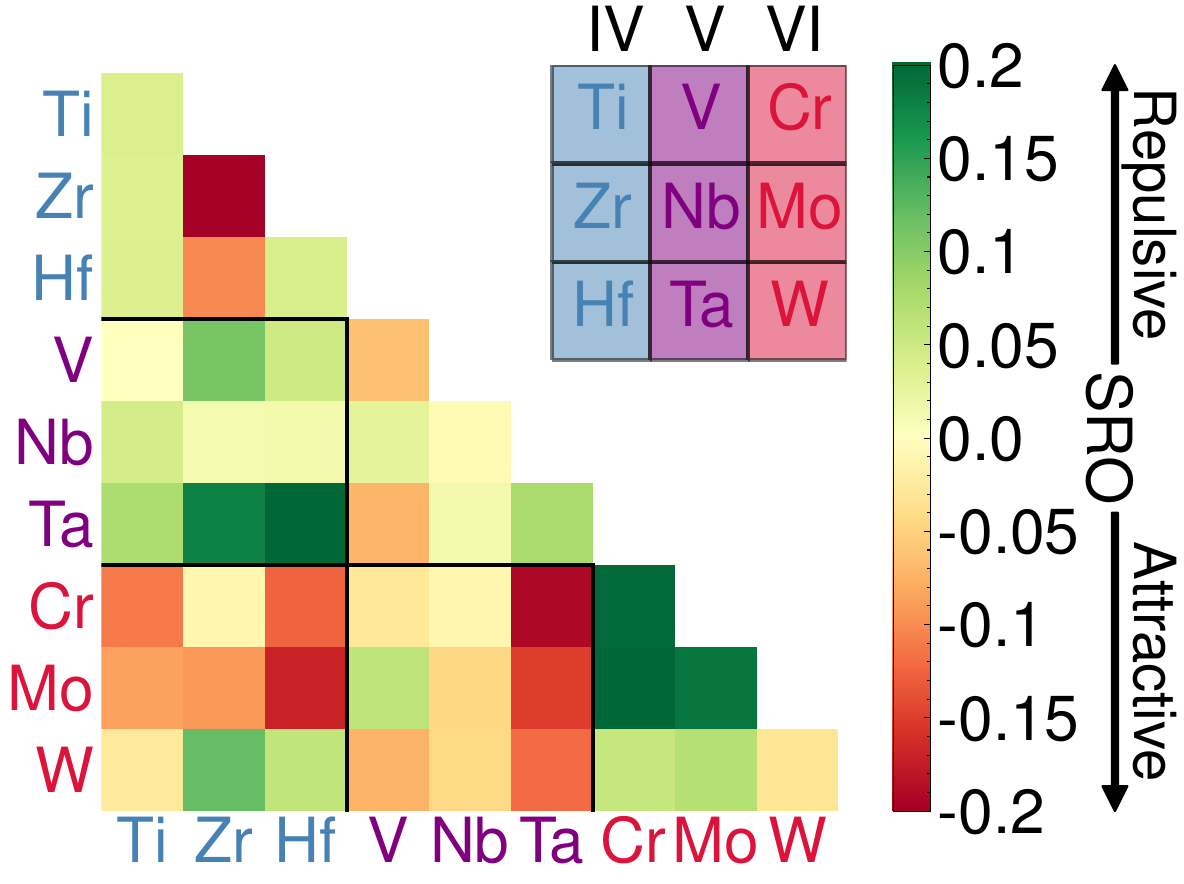}
    \caption{Warren--Cowley short-range order parameters for all element pairs in the equiatomic 9-component alloy at 2200\,K. Red (negative) values indicate attraction between the two species and green (positive) values indicate repulsion. Elements are ordered by periodic-table group, with black boxes outlining the group 4, 5, and 6 blocks. The inset shows the arrangement of the 9 elements by group and period.}
    \label{fig:9comp_sro}
  \end{figure}

  \Cref{fig:9comp_sro} shows the Warren--Cowley short-range order parameters \cite{norman_ray_1951, cowley_sro_1965, rao2022a} for all pairs in the equiatomic 9-component alloy at 2200\,K. A negative value indicates that the two species are found as neighbors more frequently than in a random alloy. Elements of group 4 tend to be attracted to those of group 6, while being repelled by group 5 elements. SRO between elements of the same group, specifically groups 5 and 6, is positive, indicating that these elements do not cluster with others of the same group, consistent with previous work \cite{muller2025}.

\begin{figure}[!ht]
    \centering
    \includegraphics[width=0.99\linewidth]{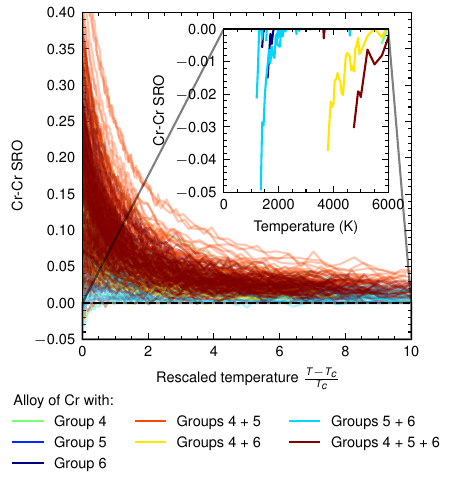}
    \caption{Warren--Cowley Cr-Cr SRO parameter for all Cr-containing equiatomic alloys in the 9-component space as a function of the rescaled temperature $(T-T_c)/T_c$, where $T_c$ is the order-disorder transition temperature of each alloy. Curves are colored by the periodic-table groups of the elements alloyed with Cr. Negative values indicate Cr clustering. The inset shows the alloys with negative Cr-Cr SRO as a function of absolute temperature.}
    \label{fig:Cr-Cr_sro}
\end{figure}

SRO can be used to probe the corrosion resistance of multicomponent alloys. Clustering of Cr atoms in the disordered state, corresponding to negative Cr-Cr SRO, has been shown \cite{xie_percolation_2021} to enable passivating oxide films to form a continuous oxide scale, addressing a major challenge facing refractory alloys. Achieving Cr clustering is difficult. The trends in \cref{fig:9comp_sro} and the results of \citeauthor{muller2025}~\cite{muller2025} both show fewer Cr-Cr pairs than in a perfectly random alloy. \Cref{fig:Cr-Cr_sro} shows the Cr-Cr SRO across all equiatomic alloys in the 9-component space, plotted for temperatures above the order-disorder transition temperature of each alloy. We identified this transition temperature from the divergence in the heat capacity computed from canonical Monte Carlo simulations. 
Most refractory alloys show positive Cr-Cr SRO, indicating that Cr atoms repel each other. We found, however, that only 5 Cr-containing alloys display attractive Cr-Cr SRO, namely CrMo, CrMoV, CrMoZr, CrMoVZr, and CrMoVW (inset, \cref{fig:Cr-Cr_sro}). Among these, CrMo and CrMoV maintain attractive SRO over a wide temperature range below 3000\,K. The disordered solid solution in the Zr-containing alloys is stable only above 3500\,K, and CrMoVW shows negative Cr-Cr SRO only within 100\,K above the critical temperature. These results highlight the role of Mo in promoting Cr clustering and, potentially, passivating oxide formation in refractory alloys, and demonstrate the utility of eCE models for rapidly screening alloy compositions for target thermodynamic properties.

\subsection{Hydrogen dissolution in refractory alloys}

Multicomponent alloys are attractive candidates for stationary hydrogen storage~\cite{schlapbach2001hydrogen, marques2021, sahlberg2016superior} and for the refractory, plasma-facing components of fusion reactors, where hydrogen-isotope retention and transport remain open concerns~\cite{zhang2026hydrogen}. In both settings, hydrogen dissolution and the underlying hydrogen-metal interactions govern performance and determine which local environments hydrogen occupies. Here we study hydrogen dissolution in a ternary Mo--Nb--W alloy. Its elements belong to groups 5 and 6 of the periodic table, and their contrasting affinities for hydrogen make the system an illustrative example~\cite{fukai2005metal}.

We model the alloy with disorder over two sublattices of a parent BCC crystal structure. The sites of the BCC host are occupied by Mo, Nb, or W, and the tetrahedral interstitial sublattice by either hydrogen or a vacancy. We assign the two sublattices different embedding dimensions, treating the ternary metal sublattice as a pseudo-binary system and applying no embedding on the interstitial sublattice.

The training dataset contains 1703 symmetrically distinct arrangements of Mo, Nb, W, and H over the two sublattices of the parent crystal structure. It includes unary, binary, and ternary decorations of the metal sublattice in supercells of up to 12 metal atoms. Hydrogen-bearing orderings, with a hydrogen-to-metal ratio (H/M) of up to 2.0, are enumerated in supercells of up to 5 metal atoms. A further 72 configurations use large supercells of 54 metal atoms and capture dilute hydrogen in a disordered metal, namely hydrogen pairs in the pure metals together with binary and ternary SQS each containing a single hydrogen atom. Together these arrangements span the quaternary Mo--Nb--W--H composition space.

The energy of each configuration was computed with DFT as implemented in the Vienna Ab initio Simulation Package (VASP)~\cite{kresse1993, kresse1996}, using the same computational settings as in previous studies~\cite{muller2025, lee2026}. The atomic positions and lattice parameters of orderings in small supercells were fully relaxed. In the 54-metal-atom supercells, only the atomic positions were relaxed, with the lattice parameters held fixed at those of the corresponding hydrogen-free metal composition.

We parameterized an eCE model with empty, point, and pair clusters up to a cutoff radius of 8\,\AA, using Chebyshev site-basis functions on the metal sublattice~\cite{vandewalle2009} and an occupation basis on the interstitial sublattice, and a feedforward neural network with three hidden layers of 32, 32, and 8 nodes. We partitioned the 1703 configurations into training (1362), validation (177), and test (164) sets. The trained model reaches mean absolute errors of 3.2, 9.4, and 10.9\,meV per metal atom on the training, validation, and held-out test sets, respectively.

Experiments measure the amount of hydrogen dissolved in a metal through the pressure-composition isotherm. Since the partial pressure of hydrogen can be related to its chemical potential $\mu_H$, an analogous isotherm can be computed with semi-grand canonical Monte Carlo simulations that vary $\mu_H$ at fixed metal composition and temperature.
Because slow diffusion keeps the metal atoms immobile, we froze the metal sublattice into an arrangement representative of the high-temperature configuration retained during synthesis. For each alloy, a snapshot of the metal ordering was drawn from canonical Monte Carlo simulations at 2000\,K. This temperature lies above the order--disorder transition temperatures of the Mo--Nb--W alloys, ensuring that the frozen host represents the disordered solid solution. The simulations used a 9$\times$9$\times$9 supercell of the conventional bcc crystal containing 1458 metal atoms and 8748 tetrahedral interstitial sites. Holding the metal configuration fixed, we then varied $\mu_H$ at 300\,K while allowing hydrogen to enter and leave the interstitial sublattice freely. At each chemical potential, the Monte Carlo simulation was continued until the 95\% confidence interval of the energy fell below 1\,meV per metal atom. Ensemble averages were computed from at least 100 uncorrelated samples.

\begin{figure}[!ht]
    \centering
    \includegraphics[width=0.99\linewidth]{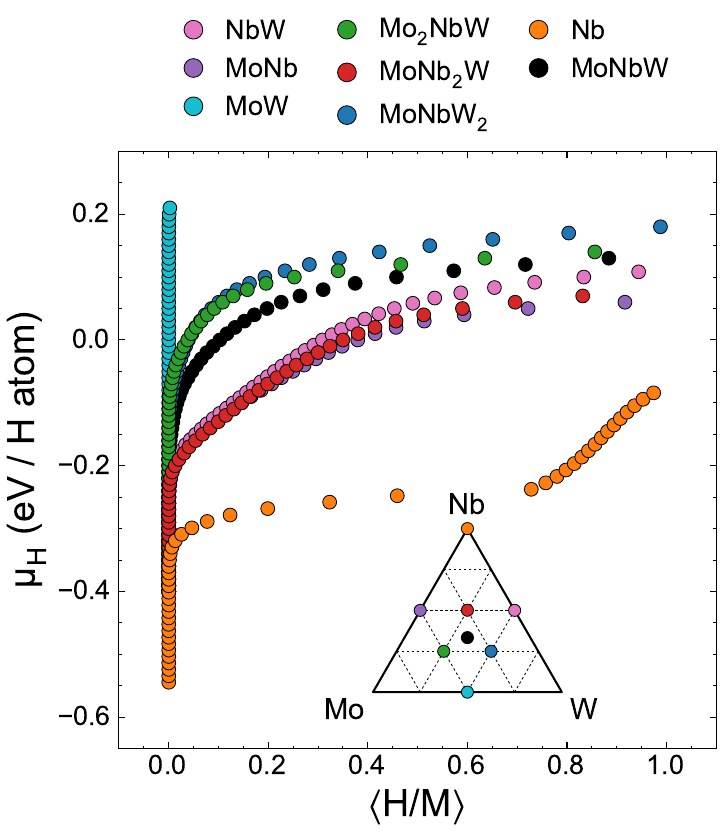}
    \caption{Computed variation in the ensemble-averaged hydrogen-to-metal ratio $\langle H/M \rangle$ with hydrogen chemical potential $\mu_H$ at 300\,K for representative Mo--Nb--W alloys, color-coded by alloy composition. The metal sublattice was held fixed in an arrangement drawn from Monte Carlo simulations at 2000\,K while hydrogen was allowed to enter and leave the interstitial sublattice.}
    \label{fig:hydrides_mu}
\end{figure}

\begin{figure}[!htbp]
    \centering
    \includegraphics[width=0.84\linewidth]{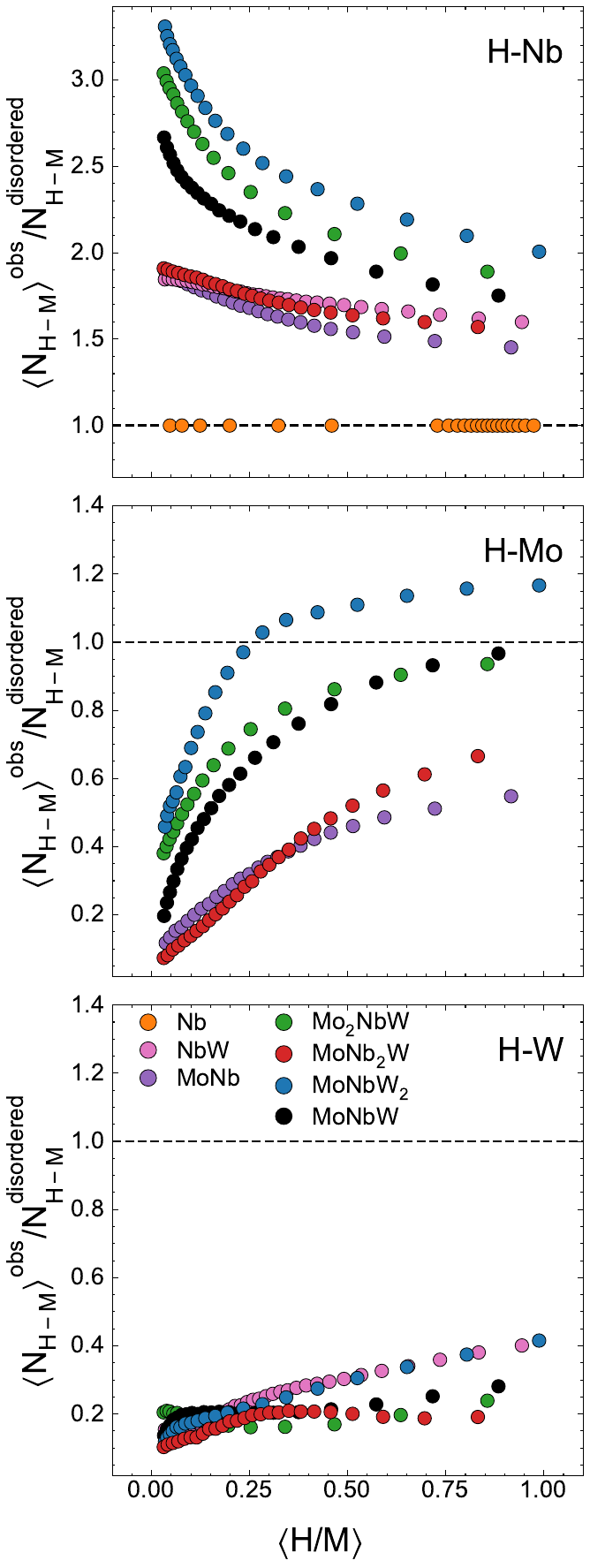}
    \caption{Ratio of the number of nearest-neighbor hydrogen-metal pairs, $\langle N^{obs}_{H-M} \rangle$, to the number expected for a random hydrogen distribution, $N^{disordered}_{H-M}$, as a function of the hydrogen content $\langle H/M \rangle$ at 300\,K. The three panels separate the H-Nb, H-Mo, and H-W interactions, with curves color-coded by alloy composition. Ratios above the dashed reference indicate preferred coordination, below it a disfavored environment. As in \cref{fig:hydrides_mu}, the metal sublattice was held fixed while hydrogen was allowed to enter and leave the interstitial sublattice.}
    \label{fig:hydrides_pairs}
\end{figure}

\Cref{fig:hydrides_mu} shows how alloy chemistry controls the amount of hydrogen dissolved at 300\,K. At fixed $\mu_H$, pure Nb dissolves the most hydrogen, Nb-rich alloys dissolve substantially more than Nb-lean alloys, and W-rich alloys dissolve the least. These trends track the stronger affinity of the group 5 element Nb for hydrogen relative to the group 6 elements Mo and W. The precise identity of the group 6 element has little effect on hydrogen solubility. For instance, replacing about half of the Mo in equiatomic MoNb with W, giving MoNb$_2$W, leaves the dissolved hydrogen almost unchanged.

In a multicomponent alloy, the metal atoms forming the tetrahedron around a dissolved hydrogen atom are likely to interact strongly with hydrogen. Consequently, environments rich in metals that interact favorably with hydrogen, such as Nb, should be preferentially populated relative to environments containing W. To assess this, we counted the nearest-neighbor hydrogen-metal pairs and compared their number with the value expected if hydrogen occupied the interstitial sites at random. A ratio of one corresponds to a random distribution, a value greater than one to preferred coordination, and a value less than one to a disfavored environment.

\Cref{fig:hydrides_pairs} shows the ratio, $\langle N^{obs}_{H-M}\rangle/N^{disordered}_{H-M}$, for each hydrogen-metal pair across the representative alloys. In every multicomponent alloy, hydrogen preferentially segregates to Nb-containing environments and occupies fewer W-containing environments than in a random distribution. The number of H-Mo pairs is also below the random value in all but MoNbW$_2$. Hydrogen therefore prefers Nb to Mo, and Mo to W, the order expected from the stronger affinity of the group 5 element. The H-W ratio is the most uniform across compositions, staying low in every alloy and rising only slightly as the W fraction grows and hydrogen can no longer avoid W neighbors. The Nb-rich ternary MoNb$_2$W and the equiatomic binaries MoNb and NbW behave almost identically in both figures (\cref{fig:hydrides_mu,fig:hydrides_pairs}). All three contain the same Nb fraction, and because both group 6 elements are unfavorable neighbors, exchanging Mo for W leaves the Nb-coordinated environments unchanged.

\section{Conclusion and Outlook}
\label{sec:outlook-conclusion}

We introduced \textit{pyeCE}, a Python package that implements the embedded cluster expansion formalism and provides a complete workflow for building, training, and deploying on-lattice models of multicomponent alloys. From a primitive cell, \textit{pyeCE} assembles a symmetry-adapted eCE model for systems with any number of species distributed over one or more sublattices, and learns a per-sublattice chemical embedding that adapts to the alloy chemistry. The site energy model is parameterized by a neural network, and \textit{pyeCE} quantifies the uncertainty of its predictions. Trained models are coarse-grained with rigorous statistical-mechanics methods to obtain finite-temperature thermodynamic properties. Built on \textit{pymatgen} and \textit{PyTorch}, it draws on established optimization algorithms and standard machine-learning workflows, and runs on GPUs. The full pipeline is exposed through both the command-line interface and the Python API.

The central advantage of the eCE formalism is that it overcomes the rapid growth in the number of cluster functions that has long limited the conventional cluster expansion to alloys of no more than three or four chemical species. By embedding many chemical species into a small set of effective ones, eCE keeps the number of cluster functions tractable while preserving the accuracy needed to resolve formation energies, short-range order, and phase stability. The two applications presented here illustrate this capability. A single model spanning a 9-component refractory system reproduced short-range order and order--disorder behavior across the full composition space, while a model of the Mo--Nb--W--H system captured hydrogen dissolution over coupled metal and interstitial sublattices. Together, these examples show that accurate, first-principles-based thermodynamic modeling of high-entropy alloys is now within reach.

The modular design of \textit{pyeCE} opens several directions for future work. The same infrastructure can support richer on-lattice models that couple chemical order to atomic displacements, magnetism, and lattice strain. It can also accommodate custom Monte Carlo schemes such as biased-ensemble sampling. The framework extends to properties beyond bulk thermodynamics, including migration barriers for diffusion, surface energetics, and the unstable stacking-fault energies that characterize planar defects. These extensions would connect complex alloy chemistries to the mechanical response that governs high-strength structural applications and to functional properties such as catalysis and energy storage. They would also enable a quantitative test of long-hypothesized high-entropy-alloy phenomena, sluggish diffusion among them, that have proven difficult to model. By coupling an interpretable chemical embedding with GPU-accelerated sampling, \textit{pyeCE} bridges atomistic calculations and finite-temperature properties, and establishes a foundation for the computational design of high-entropy alloys.

\section{Code availability}
The source code is publicly available at \url{https://github.com/epfl-mades/pyece}, and the package is distributed through \texttt{PyPI} at \url{https://pypi.org/project/pyece-mades/}.

\section{Acknowledgments}
This research was supported by the Swiss National Science Foundation through the NCCR MARVEL, a National Centre of Competence in Research (grant number 205602), and through grant number 215178. The authors also acknowledge access to Eiger at the Swiss National Supercomputing Centre under project ID mr30, allocated from the NCCR MARVEL share.

\end{document}